\documentclass[conference]{IEEEtran}
\IEEEoverridecommandlockouts
\usepackage{cite}
\usepackage{amsmath,amssymb,amsfonts}
\usepackage{algorithmic}
\usepackage{graphicx}
\usepackage{textcomp}
\usepackage{xcolor}
\usepackage{todonotes}
\usepackage{hyperref}
\usepackage{cleveref}
\usepackage{makecell}
\usepackage{braket}

\def\BibTeX{{\rm B\kern-.05em{\sc i\kern-.025em b}\kern-.08em
    T\kern-.1667em\lower.7ex\hbox{E}\kern-.125emX}}

\crefname{figure}{Fig.}{Figs.}
\Crefname{figure}{Fig.}{Figs.}
\crefname{table}{Table}{Tables}
\Crefname{table}{Table}{Tables}
\crefname{equation}{}{}
\Crefname{equation}{Equation}{Equations}
\crefname{section}{}{}
\Crefname{section}{Section}{Sections}

\title{Quantum Optimization in Wireless Communication Systems: Principles and Applications}
\author{Ioannis~Krikidis,~\IEEEmembership{Fellow,~IEEE}, and Valentin Gilbert
\thanks{I. Krikidis is with the Department of Electrical and Computer Engineering, University of Cyprus, 1678 Nicosia, Cyprus (e-mail: krikidis@ucy.ac.cy).}
\thanks{V. Gilbert is with Qounselor, 92130 Issy-les-Moulineaux, France (e-mail: valentin.gilbert@qounselor.fr).}
\thanks{This project has been supported by the European Research Council (ERC)
under the European Union’s Horizon Europe research and innovation programme,
grant agreement No. 101241675 (ERC PoC QUARTO).}}

\begin{document}
\maketitle

\begin{abstract}
Quantum optimization is poised to play a transformative role in the design of next-generation wireless communication systems by addressing key computational and technological challenges. This paper provides an overview of the principles of adiabatic quantum computing, the foundation of quantum optimization, and explores its two primary computational models: quantum annealing and the gate-based quantum approximate optimization algorithm. By highlighting their core features, performance benefits, limitations, and distinctions, we position these methods as promising tools for advancing wireless communication system design. As a case study, we examine the design of passive reconfigurable intelligent surface beamforming with binary phase-shift resolution, supported by experimental results obtained from real-world quantum hardware.
\end{abstract}
\vspace{-0.05cm}

\vspace{-0.2cm}
\section{Introduction}

\IEEEPARstart{M}{o}dern communication systems face unprecedented engineering challenges, including the need for data rates up to $1$ Tbps, massive device connectivity, ultra-reliability, energy efficiency, and air-interface latencies up to $10\times$ lower than those in current 5G deployments \cite{TAT}. To meet these demanding requirements, industry and academia are actively investigating emerging technologies and communication paradigms, such as THz frequency bands, integrated sensing and communications, advanced modulation techniques, low-bit-resolution signal processing, etc. Many of these technologies inherently involve solving complex, NP-hard combinatorial optimization problems. Traditional algorithmic approaches, including heuristics and approximation algorithms, often fall short when applied to such problems at scale, especially under real-time constraints. Furthermore, as classical (silicon-based) computing architectures approach their physical and performance limits, maintaining progress according to Moore’s Law has become increasingly reliant on architectural workarounds, such as 3D chip structures and vertical stacking. These limitations highlight the need for alternative computational paradigms \cite{KIM}.

Quantum computing has emerged as a promising alternative to this bottleneck, offering a fundamentally different approach to computation by exploiting quantum mechanical principles such as superposition, entanglement, and quantum tunneling. Two primary models dominate the field: gate-based quantum computing and quantum annealing (QA) \cite{KIM,HANZO}. The gate-based model operates through sequences of discrete, programmable logic gates acting on quantum bits (qubits). These gates perform unitary and reversible transformations, allowing for the implementation of powerful quantum algorithms offering significant theoretical speedups over classical counterparts for specific problem classes. For quantum optimization, quantum approximate optimization algorithm (QAOA) is a gate-based algorithm tailored for near-term quantum hardware. It operates in a hybrid quantum-classical loop, alternating between applying parameterized quantum circuits and classical optimization routines to approximate the solution of a combinatorial problem. Due to its adaptability and compatibility with noisy intermediate-scale quantum (NISQ) devices, QAOA is increasingly seen as a practical approach for early-stage quantum advantage in real-world applications.

In contrast, QA is an analog model rooted in the adiabatic principles of quantum mechanics. QA is particularly well-suited for solving NP-hard combinatorial optimization problems, often formulated as quadratic unconstrained binary optimization (QUBO) problems or Ising models. It operates by gradually evolving a quantum system from an initial ground state (state with the lowest energy) to the ground state of a final Hamiltonian (energy function) that encodes the objective function of the optimization problem. Commercial systems like D-Wave have implemented QA using superconducting qubit architectures and have demonstrated its effectiveness in approximating large-scale NP-hard problems, including various applications in wireless communications.

More broadly, one of the most promising application domains for QA/QAOA is operations research, where tasks such as logistics, supply chain management, network routing, and vehicle scheduling involve combinatorial spaces that quickly become intractable. These problems are typically expressed as cost functions $f(x)$, with the goal of identifying the configuration $x$ that optimizes the objective. Quantum optimization has also been applied to wireless communications, with early work focusing on ML detection in large-scale MIMO systems and showing encouraging gains despite hardware limitations. More recent efforts include beam assignment, user scheduling, RIS phase-shift design, multi-user detection, polar code decoding, and antenna selection in reconfigurable MIMO systems \cite{KASI, KRI1, LIM, zhao2024quantum}. Although current quantum systems remain costly and limited due to cryogenic operation and specialized environments, advances in qubit scalability, connectivity, and error mitigation are rapidly improving accessibility. In practical architectures, quantum processors could be integrated into centralized networks, where base stations offload heavy optimization workloads such as ML-MIMO detection \cite{KIM}, user scheduling, and RIS configuration. Thus, quantum optimization is expected to complement rather than replace classical methods, acting as an accelerator for complex wireless design challenges -an emerging direction of growing relevance.

In this tutorial paper, we introduce the concept of quantum optimization within the context of communication engineering and investigate the potential of both QA and gate-based quantum computing for approximating solutions to combinatorial optimization problems. We begin by outlining the principles of adiabatic quantum computing, which underlie quantum optimization, and present its two primary models: QA with focus on the D-Wave implementation, and the gate-based QAOA, which serves as a Trotterized version of adiabatic computing. We further discuss the fundamental and technological limitations of both approaches, along with open research challenges. At this stage, quantum methods should be regarded as promising heuristics rather than proven outperformers of classical algorithms; our contribution is to introduce these concepts and demonstrate their application to wireless communication use cases. As an illustrative case study, we consider the design of passive beamforming for a RIS with binary phase-shift resolution. Practical implementations are demonstrated using state-of-the-art D-Wave systems for QA and quantum circuits executed on IBM quantum devices for QAOA. Although classical algorithms can efficiently solve the small-scale problem considered here, this example serves to illustrate how quantum solvers operate in practice and to highlight their potential for larger, computationally intractable wireless optimization tasks.

\vspace{-0.1cm}
\section{Quantum optimization: from analog quantum computing to gate-based model}

In this section, we introduce the fundamental concept of adiabatic evolution, which underpins quantum optimization. We also present a noisy, practical implementation known as QA, along with its discretized gate-based counterpart {\it i.e.,} QAOA.
\vspace{-0.1cm}
\subsection{Adiabatic and analog quantum computing}

The exploitation of quantum fluctuations to accelerate the convergence of classical optimization methods such as simulated annealing (SA) was first exposed by T. Kadowaki et al. in \cite{kadowaki1998quantum}. While SA relies on thermal fluctuations and a cooling schedule to probabilistically escape local minima, QA replaces thermal fluctuations with quantum fluctuations, enabling the system to escape local minima through quantum tunneling. This fundamental difference allows QA, in principle, to overcome energy barriers that may trap classical heuristics. Building on this idea, the quantum adiabatic algorithm was introduced by E. Farhi et al. \cite{farhi2000quantum} to address optimization problems with quantum computers.

The resolution of the problem relies on a theorem that guarantees the convergence of the quantum algorithm, considering that the system is prepared in a specific initial state and that the evolution is perfectly controlled without noise. Such ideal systems are usually called \textit{closed} quantum systems, in opposition to \textit{open} systems for noisy devices. The evolution of a quantum system can be described by an operator $H$ called the \textit{Hamiltonian} of the system; it corresponds to a matrix with complex coefficients. If the operator changes through time, the Hamiltonian $H(s)$ characterizes the evolution at each annealing fraction $s=t/t_\mathrm{f}$ where $t_\mathrm{f}$ corresponds to the final annealing runtime and $t \in [0, t_\mathrm{f}]$. The eigendecomposition of the matrix $H(s)$ at each time step $s$ gives the list of eigenvalues (\textit{eigenenergies}). An eigenvector (\textit{eigenstate}) associated to the lowest eigenvalue is usually called a \textit{ground state}. The ground state is said to be \textit{degenerate}, if more than one state is associated with the lowest eigenvalue. The \textit{adiabatic theorem} ensures that a quantum system initialized in the ground state of some Hamiltonian $H(0)$ will remain in the instantaneous ground state of the Hamiltonian at each annealing fraction $s$ given that the system's evolution is slow enough ({\it i.e.,} $t_\mathrm{f}$ is large enough). \\
The idea of quantum manufacturers was to leverage this theorem to build analog-based quantum computers with a time-dependent Hamiltonian based on the interpolation of two time-independent Hamiltonians: an initial one $H_\mathrm{I}$ and a final one $H_\mathrm{F}$. This interpolation is noted as
\begin{equation}
    H(s) = -A(s) H_\mathrm{I} + B(s) H_\mathrm{F}.
\end{equation}

The functions $A(s)$ and $B(s)$ are called the \textit{annealing scheduling functions} and mainly depend on the implementation. At the beginning of the evolution, the system is prepared in the ground state of $H_\mathrm{I}$, which dominates the evolution ($A(0) \gg B(0)$). $H_\mathrm{I}$ is often imposed by the manufacturer to ease the preparation of the ground state of this Hamiltonian. In the case of D-Wave systems, $H_\mathrm{I}$ is chosen such that its ground state is a uniform superposition of the states of the computational basis. When time passes, $A(s)$ is slowly turned off, and $B(s)$ is being activated. At the end of the evolution, $B(1) \gg A(1)$ such that the system finishes in the ground state of $H_\mathrm{F}$. The final Hamiltonian $H_\mathrm{F}$ can be programmed and is used to encode the optimization problem where the ground state corresponds to the optimal solution of the problem. This correspondence permits the extraction of the solution to the optimization problem when measuring the quantum state.

\begin{figure}[t!]
    \centering
    \includegraphics[width=\columnwidth]{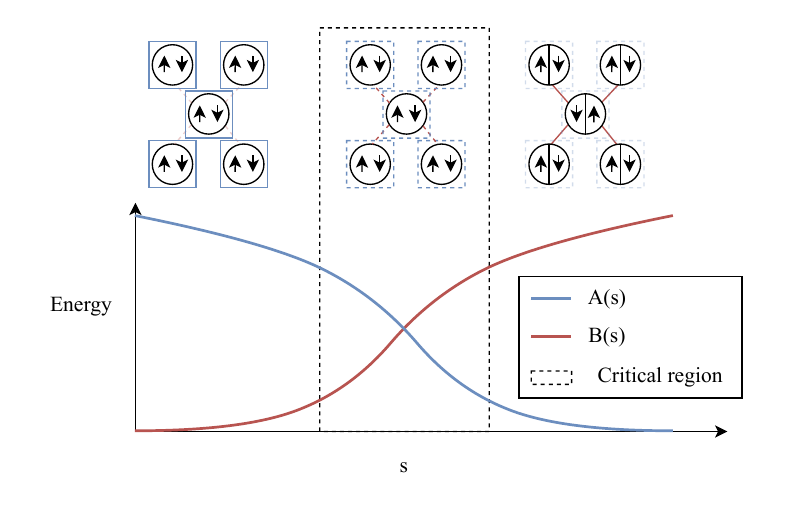}
     \vspace{-1cm}
    \caption[]{QA schedule. The lower part of the figure shows examples of annealing schedule functions $A(s)$ and $B(s)$. $A(s)$ (resp. $B(s)$) drives the strength of the magnetic field of the initial Hamiltonian $H_\mathrm{I}$ (resp. the final Hamiltonian $H_\mathrm{F}$). The upper part of the figure shows how the qubits and couplers are affected by the magnetic field related to each Hamiltonian. At the beginning of the evolution, the system is in a disordered phase with all the qubits in superposition. At the end of the evolution, the system is in an ordered phase where qubits alignment is fixed. Between these two phases, the system goes through a phase transition likely to appear in the critical region when $A(s) \approx B(s)$. The precise location of the phase transition depends on the shape of the schedule and on the Hamiltonian. At the end of the evolution, measuring the qubits results in either the left half or the right half of each circle (the state being measured is in a superposition of the two optimal solutions to the MaxCut problem).
    }
    \label{fig:quantum_annealing}
\end{figure}
To further demonstrate the adiabatic evolution, we show a toy instance of the MaxCut problem\footnote{The problem involves assigning labels to nodes such that an edge is considered a cut, if it connects two nodes with different labels. The objective is to maximize the total number of such cuts. In our example, the graph consists of five nodes arranged in a star topology: one central node connected to four peripheral nodes, which are not connected to each other.} in \cref{fig:quantum_annealing}. At the beginning of the evolution, the label of each node is in a superposition of up and down states, which corresponds to the ground state of $H_\mathrm{I}$. $H_\mathrm{I}$ is then slowly turned off, and $H_\mathrm{F}$ starts to be activated. When $A(s) \approx B(s)$, the system enters a critical region where the qubits are influenced both by $H_\mathrm{I}$ and $H_\mathrm{F}$. At some point, when $H_\mathrm{F}$ will dominate the evolution, the quantum state converges the ground state of $H_\mathrm{F}$. The result of the measurement of this state gives the solution to the optimization problem.

An important question in adiabatic quantum computing is how to determine whether the quantum evolution is sufficiently slow to ensure reliable solutions. The evolution time is fixed by the \textit{spectral gap} $\Delta_\mathrm{min}$ and depends on the problem being solved. Specifically, at each annealing fraction $s$, it is possible to compute the full spectrum of the eigenvalues of the Hamiltonian $H(s)$. The spectral gap is the smallest gap of energy between the ground state of the system (state of lowest energy) and the first excited state at any annealing fraction $s$. To evolve according to the adiabatic theorem, the time $t_\mathrm{f}$ should at least scale as $O(1/\Delta_\mathrm{min}^2)$ \cite{albash2018adiabatic}. In practice, computing this gap is at least as difficult as solving the optimization problem itself. It has been shown that some problem instances exhibit exponentially small gaps with the scaling of the problem size, meaning that adiabatic quantum computing can not bring any advantage compared to classical approaches for these instances. 

\begin{figure}[t!]
    \centering
    \includegraphics[width=\columnwidth]{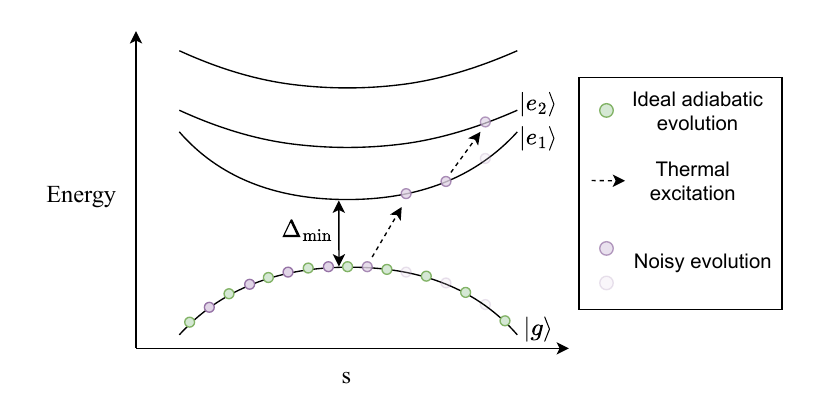}
     \vspace{-1cm}
    \caption[]{Eigenenergy diagram of the schedule-dependent Hamiltonian $H(s)$ with a unique ground state. The quantum system is initialized in the ground state of the initial Hamiltonian $H_\mathrm{I}$. The qubits remain in the ground state during the whole evolution for closed systems (green dots). In case of noisy interaction, the quantum state might suffer from thermal excitations and become excited. The probability of observing the ground state diminishes, and the quantum state might finish in a superposition of several excited states (purple dots where transparent dots mean lower probability of observation). Passing through the critical region too rapidly also reduces the probability of observing the ground state at the end of the evolution.}
    \label{fig:eigenstate_energies}
\end{figure}

QAs are noisy implementations of the adiabatic theorem, often featuring imperfect qubit initialization and limited control over the Hamiltonians $H_\mathrm{I}$ and $H_\mathrm{F}$. In such quantum chips, various factors can disturb the evolution and result in the measurement of excited states rather than the ground state. For instance, qubits may experience unwanted interactions with the environment that can excite the system. In such cases, the probability of measuring the ground state at the end of the annealing process becomes vanishingly small (see \cref{fig:eigenstate_energies}).

D-Wave (British Columbia, Canada) \cite{WAVE}, is a pioneer company in quantum computing and has developed a line of commercial quantum processors that implement QA. The largest D-Wave system features over $5,500$ qubits, making it one of the largest QA devices available today. These qubits are connected in a specialized hardware layout (known as the Pegasus topology), where each qubit can connect to up to $15$ others; this high connectivity allows us to represent large and complex optimization problems. The first step to map a combinatorial problem onto D-Wave is to express the optimization problem’s objective function as an \textit{Ising model} cost function with binary spin variables $s_i = \pm 1$, given by
\begin{equation}
    C(\mathbf{s}) = \sum_{i=1}^n h_i s_i + \sum_{i<j} J_{i,j} s_i s_j,
    \label{eqn:ising_cost_function}
\end{equation}
where $\mathbf{s} = (s_1, s_2, ..., s_n)$, while the coefficients $h_i$ (linear terms) and $J_{i,j}$ (quadratic couplings) define the energy landscape of the system. These parameters are used to program the final Hamiltonian $H_\mathrm{F}$, whose ground state corresponds to the configuration that minimizes the cost function $C(\mathbf{s})$. As an illustrative example, consider the toy MaxCut instance in Fig. \ref{fig:quantum_annealing}, where the associated Ising cost function to minimize is given by $C(\mathbf{s}) = s_1 s_3 + s_2 s_3 + s_4 s_3 + s_5 s_3$.

The second step is to map the cost function onto the quantum chip, which is represented as a hardware graph, where the vertices correspond to qubits and the edges to programmable couplers. Each qubit includes an auto-coupler, which is used to program the local field values $h_i$, while to program an interaction terms $J_{i,j}$, qubit $i$ must be physically connected to qubit $j$ via a coupler. Consequently, the quadratic terms in the cost function $C(\mathbf{s})$ define the qubit connectivity required to directly program the quantum Hamiltonian. Since the qubit graph of a D-Wave quantum computer is not fully connected, some qubit-to-qubit couplings may be missing when the quadratic expression is dense. In such cases, an additional step (known as \textit{embedding}) is required to map the cost function onto the hardware, often using multiple physical qubits to represent a single logical qubit. Note also that higher-order terms such as $s_i s_j s_k$ (cubic) can not be directly implemented on the quantum chip and must first be reduced to quadratic form by using auxiliary variables. Once an embedding is found, the problem can be encoded on the chip. The range of values that can be assigned to the auto-couplers and couplers is also bounded, so rescaling these coefficients to fit within the allowed range/resolution is usually necessary. This rescaling affects the spectral gap proportionally and may influence solution quality. 

D-Wave QA is also subject to noise and hardware imperfections, known as \textit{integrated control errors}, including flux noise, qubit crosstalk, finite numerical resolution, etc. Due to these non-idealities, the result of a single computation (called {\it anneal}) is probabilistic and may not always produce the best solution. To improve reliability, the same problem is usually run many times, and the best result among those runs/anneals is selected as the final solution. The key system parameters such as the annealing time and the total number of anneals must be carefully tuned for optimal performance.

\vspace{-0.1cm}
\subsection{A gate-based model: QAOA}

\begin{figure}[t!]
    \centering
    \includegraphics[width=\columnwidth]{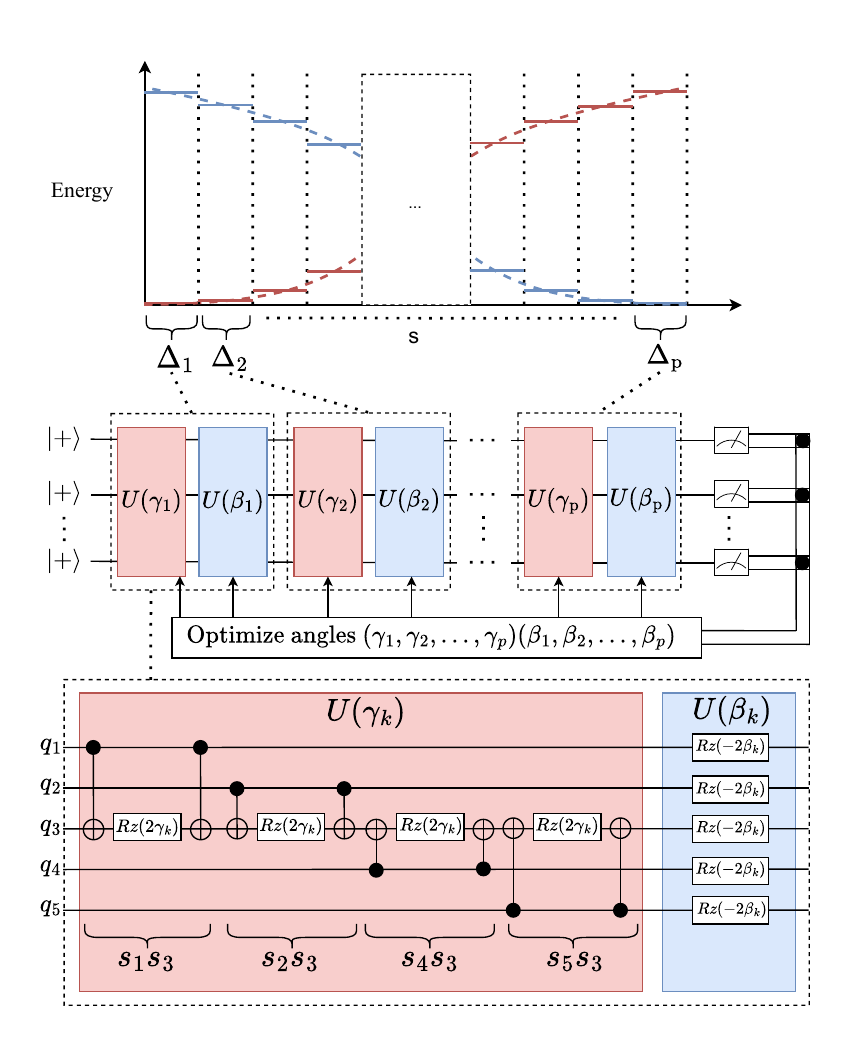}
    \vspace{-1cm}
    \caption[]{Discretization of the quantum adiabatic process in $p$ constant steps. The upper part shows a piecewise approximation of the annealing schedule. The medium part shows a quantum digital circuit where each qubit is associated with a wire. The qubits are prepared in a uniform superposition of states of the computational basis. Each constant time step $\Delta_k$ is used to build the unitary $U(\beta_k)$ (implementing the Hamiltonian $H_\mathrm{I}$) and the unitary $U(\gamma_k)$ (implementing the Hamiltonian $H_\mathrm{F}$). The schedule associated with each $H_\mathrm{I}$ and $H_\mathrm{F}$ is transmitted to the unitaries with the parameters $\beta_k$ and $\gamma_k$. The lower part displays a single layer of QAOA circuit corresponding to the MaxCut instance problem \cref{fig:quantum_annealing}. Each variable $s_i$ is associated with a qubit $q_i$. The $U(\beta_k)$ layer corresponds to a single wall of single rotation gates $R_x(-2\beta_k)$. In the layer associated to the problem $U(\gamma_k)$ each quadratic term is built from $2$ CNOTs and a single rotation gate  $R_z(2\gamma_k)$.}
    \label{fig:qaoa_circuit}
\end{figure}

The QAOA was introduced by E. Farhi et al. in 2014 \cite{farhi2014quantum}. This algorithm is based on discretizing the quantum adiabatic process in $p$ constant time steps $(\Delta_1, \Delta_2, ..., \Delta_p)$ (see \cref{fig:qaoa_circuit}). The parameter $p$ is often referred as the depth of the QAOA. When $p$ is large enough, each time-step evolution combining $H_\mathrm{I}$ and $H_\mathrm{F}$ during $\Delta_k$ time step is approximated by two unitary transformations $U(\gamma_k)$ and $U(\beta_k)$. This approximation is exact as $\Delta_k \rightarrow 0$ and is based on the Suzuki-Trotter product formula (errors introduced by this approximation process are called \textit{Trotter errors}). This approximation permits the conversion of a continuous time Hamiltonian into a product of fixed-time Hamiltonians used to build the quantum circuit with parametrized unitaries. If the cost function to minimize is an Ising cost function, the structure of the circuit is very simple: the unitary corresponding to $H_\mathrm{I}$ at each time step is composed of a single wall of parametrized rotation $R_x(-2\beta_k)$ gates\footnote{In quantum mechanics, single-qubit rotation gates such as \( R_x(\theta)\) and \( R_z(\theta)\) represent unitary operations that rotate the qubit state around the \( x \)- and \( z \)-axes of the Bloch sphere by an angle \( \theta \), respectively.}. The unitary associated to $H_\mathrm{F}$ is composed of CNOT and $R_z$ gates {\it i.e.,} linear terms $h_is_i$ in the cost function are traduced by a single gate $R_z(2h_i\gamma_k)$ on the qubit $i$, and quadratic terms $J_{i,j}s_is_j$ are translated to two CNOT gates and an $R_z(2J_{i,j}\gamma_k)$ gate on qubits $i$ and $j$. A QAOA circuit layer example associated with the MaxCut instance of \cref{fig:quantum_annealing} is shown in \cref{fig:qaoa_circuit}. The expectation value computed from the output of the parametrized circuit is used to optimize the variables $(\gamma_1, \gamma_2, ...., \gamma_p)$ and $(\beta_1, \beta_2, ...., \beta_p)$; the values for the angles $\gamma_k$ and $\beta_k$ are usually bounded by $[0, 2\pi]$. In practice, these ranges can be reduced for some specific problems that possess symmetries in the solution space (it is the case for the Maxcut problem).

Various strategies exist to optimize the angles of the QAOA scheme. A popular approach is to start at depth $p=1$ with only two angles initialized randomly. These angles are then optimized with a local search method such as a gradient descent algorithm or other numerical methods {\it e.g.,} COBYLA. The depth of the QAOA is then increased with an additional layer with two new angles initialized with an educated guess. After this step, the local search algorithm is used to optimize the four angles. This method is repeated to arbitrary depth and produces good expectation values in practice \cite{zhou2020quantum}.

\vspace{-0.3cm}
\section{Quantum optimization: Performance, limitations and challenges}

Each of the quantum optimization methods discussed above presents potential performance advantages, yet also entails fundamental limitations. In this section, we examine the key technological constraints and open challenges that must be addressed to advance the field. The limitations of QA are closely linked to the characteristics of the underlying hardware architecture, which can vary significantly across different platforms. Our focus here is on the D-Wave systems, currently the most advanced commercially available devices for implementing QA. The primary technological limitations and associated challenges of these systems are summarized below.

\noindent {\it Adiabatic intrinsic limit:} The computational time required for adiabatic quantum computing is fundamentally constrained by the size of the spectral gap {\it i.e.,} the minimum energy difference between the ground and first excited states during the evolution ($\Delta_\mathrm{min}$). However, determining this gap is at least as hard as solving the original optimization problem, making it generally inaccessible in advance. Instances in which the spectral gap decreases exponentially with problem's size pose a significant challenge for adiabatic computing, as they demand prohibitively long evolution times to satisfy the adiabatic condition.

\noindent {\it Hardware topology:} In contrast to neutral atoms quantum computers, which have reconfigurable arrays of qubits, D-Wave systems are built over superconducting qubits and possess fixed topologies with predefined qubit location and interconnection. The qubit interconnection graph is sparse, with at most $15\sim20$ connections per qubit for the most connected quantum chips of today. It is obvious that Ising cost functions with many quadratic terms rapidly go beyond this density limit. For these cases, it is necessary to find a mapping procedure to map each variable of the Ising cost function to the qubits of the hardware graph. Finding the lowest number of qubits to embed an Ising cost function is itself an NP-hard problem. Although polynomial-time techniques are available to embed arbitrarily dense graphs of size $n$ with a qubit overhead scaling as $O(n^2)$, the embedded instance generally presents greater computational difficulty than the initial problem. In addition, mapping a single variable to several qubits requires accurate setting of coupling strengths to make them act as a single (logical) qubit.

\noindent {\it System noise:} Recent studies have shown that  D-Wave systems can operate under a very short annealing time regime (annealing evolution in the order of nanosecond). In this regime, the quantum computer behaves as a closed quantum system and the spectral gap analysis permits to evaluate the convergence of the quantum computer. However, large optimization problems are expected to have a very small spectral gap, requiring a longer annealing time to converge to the optimal solution. The system's evolution becomes noisy beyond $10$ nanoseconds of annealing time and these errors violate the adiabatic theorem. Therefore, there is no guarantee of convergence of the QA at a longer annealing time (in practice, the QA can not converge to optimal solutions for large problems). 

\noindent {\it Computational advantage:} The computational advantage of D-Wave systems for solving optimization problems remains an open question. While some studies have reported scaling advantages in specific cases, there is still no strong theoretical evidence that QA consistently outperforms classical optimization solvers. Such an advantage is more likely to emerge in problem instances that are tailored to the QA, particularly when the problem's connectivity structure closely matches the hardware's qubit interconnection graph. Research aiming to demonstrate the benefits of QAs often focuses on their scaling advantage over classical heuristics, such as SA or parallel tempering, typically using annealing time as the sole metric in evaluating time to solution ({\it i.e.,} the time required to find the optimal solution). However, in practice, the annealing time represents only a small fraction of the total wall-clock time required to solve a problem on current quantum hardware, as the overall runtime is dominated by external overheads such as initialization, readout, and control latency.

As with the QA approach, the performance of the QAOA is heavily influenced by the underlying hardware platform, with each technology offering distinct trade-offs. For instance, superconducting quantum chips provide fast gate operations (on the order of nanoseconds) and promising scalability, but they are limited by short coherence times and restricted qubit connectivity. In contrast, trapped-ion systems feature long coherence times and full qubit connectivity, yet suffer from slower gate speeds (typically hundreds of nanoseconds) and face significant challenges in scaling to larger qubit counts. The key limitations and challenges associated with QAOA are summarized below.

\noindent {\it Intrinsic limit:} When the number of layers $p \rightarrow \infty$, the QAOA process is fully equivalent to an adiabatic process. If the QAOA is executed on a noisy quantum chip, it suffers the same limits as the QA. The depth of the QAOA circuit can be tuned by changing the parameter $p$. There is, therefore, a tradeoff between the approximation quality that minimizes the Trotter error and the quantum circuit depth. In noisy environments, a careful study of this tradeoff has to be considered to improve the results returned by the QAOA \cite{blekos2024review}.

\noindent {\it Hardware topology:} For superconducting qubits, current gate-based hardware generally remains too noisy to reliably execute problem instances involving more than $10$ qubits and hundreds of gate operations, effectively limiting QAOA to no more than $2$ or $3$ layers in practice. Larger instances, involving up to 18 qubits, have been successfully demonstrated on trapped-ion quantum computers. Overall, the quantum hardware performance degrades significantly when the problem instance does not align with the native qubit connectivity of the chip, requiring additional SWAP gates for execution.

\noindent {\it Angle optimization:} One fundamental limit of the QAOA is the optimization of angles that require the execution of the circuit on the quantum chip, as the optimization is based on the average expectation value of the unitary. The parameter concentration in QAOA has been demonstrated both analytically and empirically \cite{blekos2024review}, indicating that good initial angles (likely to yield high-quality solutions) can be identified and supplied to the optimizer. However, efficient training of the QAOA angles on noisy quantum computers remains challenging due to barren plateaus induced by noise.

\noindent {\it Computational universality:} One key advantage of QAOA over D-Wave's QA systems is that it is based on a \textit{universal quantum computation model}, allowing the implementation of arbitrary quantum evolutions~\cite{lloyd2018quantum}. In contrast, D-Wave systems impose hardware-specific constraints on the form of the initial and final Hamiltonians, \( H_\mathrm{I} \) and \( H_\mathrm{F} \), limiting the flexibility of the quantum evolution. With QAOA, the user can define both \( H_\mathrm{I} \) and \( H_\mathrm{F} \), enabling a broader range of problem encodings and making the approach inherently more general and programmable than D-Wave’s annealing process. QAOA provides a straightforward mechanism to encode higher-order interactions, as the required circuit depth increases only linearly with the interaction order. In contrast, QA can represent higher-order terms only indirectly, by employing quadratization strategies that expand the QUBO formulation and introduce auxiliary binary constraints. We note that many optimization problems in wireless communications ({\it e.g.,} user scheduling and resource allocation, ML MIMO with high order modulation, etc.) are subject to strict equality and inequality resource constraints, where either the objective function or the constraints may involve higher-order polynomials.

It is worth noting that, unlike classical optimization algorithms where computational complexity can often be explicitly characterized, the complexity of QA and QAOA remains largely problem-dependent and unresolved. For QA, the runtime is linked to the minimum spectral gap $\Delta_\mathrm{min}$, which is generally intractable to compute, while for QAOA, complexity is tied to circuit depth and parameter optimization overhead. Consequently, current studies focus mainly on empirical performance and scaling behavior rather than formal complexity bounds.

\begin{figure}[t!]
    \centering
    \includegraphics[width=0.9\columnwidth]{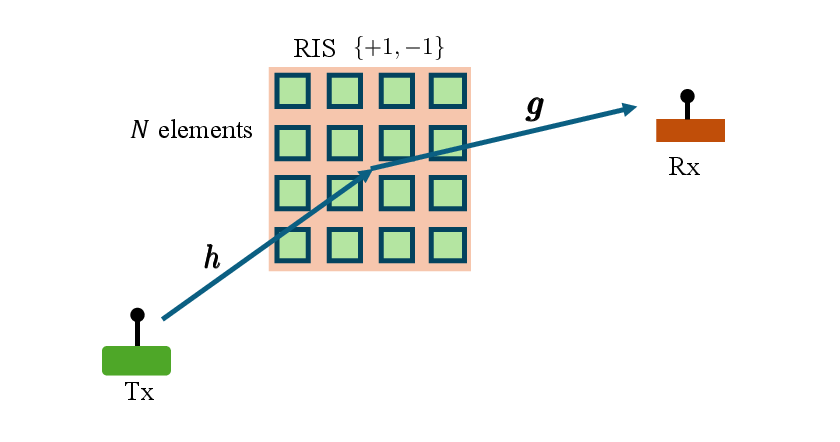}
    \vspace{-0.7cm}
    \caption[]{The system model for a RIS-aided communication system with binary phase-shift RIS resolution. }
    \label{fig:model}
\end{figure}

\vspace{-0.2cm}
\section{A case study: Passive beamforming in $1$-bit RIS}

To demonstrate the potential interest of the above quantum solvers in communication systems, we study the design of RIS passive beamforming with binary phase-shift resolution. Specifically, we consider a simple RIS-assisted wireless communication system comprising a single-antenna transmitter, a single-antenna receiver, and a passive RIS equipped with $N$ reflecting elements. The wireless channels from the transmitter to the RIS ($\pmb{h}$) and from the RIS to the receiver ($\pmb{g}$) follow specific fading distribution and are modeled as complex-valued vectors of dimension $N$, while the phase-shift RIS matrix is diagonal of dimension $N\times N$. In our setup, a direct line-of-sight path between the transmitter and receiver is absent due to significant obstacles and deep shadowing. As a result, all signal transmission is exclusively dependent on the RIS, which passively reflects the signal from the transmitter to the receiver.

To minimize power consumption and signaling/control overhead, the RIS employs binary phase shifts to the incident signal. Specifically, each reflecting element can induce a phase shift of either $0$ or $\pi$, corresponding to multiplying the incoming signal by $+1$ or $-1$, respectively. The signal-to-noise ratio (SNR) at the receiver, which characterizes the quality of the received signal, depends on the transmit power, the phase-shift configuration at the RIS, the underlying wireless channels, and the noise level. It is assumed that global channel state information is available at both the transmitter and the receiver. A schematic representation of the system is shown in Fig. \ref{fig:model}. Within this framework, the conventional beamforming approach using binary phase-shifts aims to configure the RIS in a way that maximizes the SNR at the receiver. This design challenge can be reformulated as an Ising problem (equivalent to QUBO), where the goal is to identify the optimal binary phase-shift vector that yields the highest SNR. Such problems are well-suited to quantum optimization methods and can be efficiently tackled using quantum solvers such as the D-Wave QA and QAOA-based heuristics.

\begin{figure}
    \centering
    \includegraphics[width=\columnwidth]{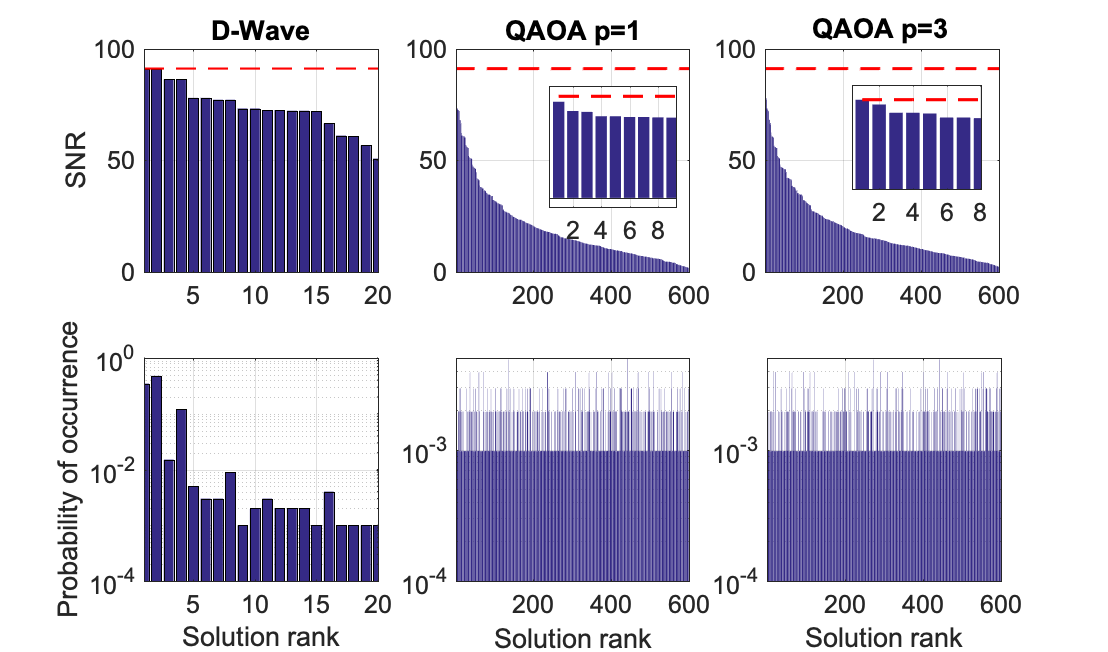}
    \vspace{-0.6cm}
    \caption[]{D-Wave versus QAOA with $p=\{1,3\}$. (top) SNR for the ordered solutions; exhaustive search (dashed line). (bottom) Probability of occurrence for the ordered solutions.}
    \label{fig:results}
\end{figure}

\vspace{-0.2cm}
\subsection{Experimental setup and results}
We consider a RIS with $N=10$ reflecting elements, the transmitted power and the noise variance are normalized to one, and the wireless channels follow a complex Gaussian distribution with unit variance (Rayleigh fading channel); we study a single random realization. The RIS size of $10$ elements serves as a simple illustrative example that can be implemented on both experimental platforms and is sufficient to capture the key trends of the two solvers. While both QA and QAOA face hardware connectivity and scalability constraints, QA is primarily limited by embedding overhead, whereas QAOA requires increasingly deep circuits as the problem size grows. In gate-based platforms, the combination of limited qubit counts, sparse connectivity, and the need for explicit error-mitigation techniques further amplifies noise sensitivity, making large-scale RIS configurations challenging for current QAOA implementations. Overall, QA hardware currently appears more mature and stable for mid-scale problems.

For our D-Wave QA experiments, we use the D-Wave Leap interface with the {\it Advantage\_System6.4} quantum processing unit \cite{WAVE}. We use $1,000$ anneals with $1\mu$sec anneal time; the chain strength (ferromagnetic coupling parameter) is tuned equal to $3$ \cite{KRI1}. For minor embedding, we adopt the heuristic algorithm {\it minorminer} which is included in the Ocean SDK by default, with majority vote decoding. The QAOA circuit (with depth $p={1,3}$) is executed on the IBM Quantum backend {\it ibm\_sherbrooke} using Qiskit SDK. Circuit parameters are initialized manually and optimized with the COBYLA algorithm (maximum $300$ iterations). For the three-layer QAOA, all parameters are optimized from scratch without any educated guess or warm start. The parameter optimization is performed entirely in simulation using Qiskit's {\it AerSimulator}, where the circuit is repeatedly evaluated to minimize the cost function. After the optimal parameters are obtained, the final circuit is transpiled with optimization level $1$ (basic techniques to reduce gate count and adjust layout) and executed on real quantum hardware via {\it SamplerV2}, using $1024$ independent shots. Error mitigation techniques, including {\it XY4} dynamical decoupling and gate twirling, are enabled during execution. Measurement outcomes are collected for cost evaluation and solution analysis.

Figure \ref{fig:results} illustrates the performance of the two solvers in terms of solution quality. Specifically, it shows the objective function values (SNR) for the ordered solutions returned by each solver, along with their corresponding probabilities of occurrence. The results from exhaustive search\footnote{Exhaustive search is used as a sufficient performance benchmark, while comparison with other traditional heuristics (e.g., SA, genetic algorithms) is beyond the scope of this work.} (representing the optimal solution) are included as a benchmark. Both solvers exhibit near-optimal performance. However, a closer examination reveals that D-Wave tends to produce higher-quality solutions, characterized by a smaller set of distinct outcomes (only $20$ solutions) and a higher probability assigned to the optimal ones. In contrast, the QAOA technique, while improving as the number of layers increases, yields a significantly larger number of distinct solutions. Moreover, the probability associated with its best solutions remains notably lower than that achieved by the D-Wave heuristic.

\vspace{-0.3cm}
\section{Conclusion}

The integration of quantum computing tools, specifically quantum optimization, into the design of wireless communication systems is an emerging research topic. Given the extremely high computational demands of next-generation communication systems, quantum (non-conventional) computing machines are expected to play a pivotal role in their design and orchestration. In this paper, we focus on quantum optimization techniques and introduce two fundamental quantum solvers/models {\it i.e.,} QA (with emphasis on the D-Wave) and QAOA. Although both heuristics have appeared in the communication theory literature to solve various problems, their underlying principles and technical details have often been overlooked. We provide a thorough discussion of their relationship to analog adiabatic computing, key characteristics, differences, limitations, and open challenges. As a case study, we apply both quantum optimization models to the design of RIS passive beamforming with binary phase-shift resolution, demonstrating experimental results using real-world quantum devices. Our results indicate that, for the considered setup, D-Wave QA tends to provide higher-quality solutions than the tested QAOA implementation, reflecting the relative maturity of current annealing hardware. Further integration of quantum optimization into wireless communication system design will require significant theoretical and technological advancements, not only in quantum engineering (to ensure more reliable, robust, and scalable quantum devices), but also in communication theory, to identify problems and techniques that can exploit these new computing paradigms \cite{zhao2024quantum}.

\vspace{-0.1cm}
\bibliographystyle{ieeetr}
\bibliography{bibliography}

\vspace{-1.2cm}
\begin{IEEEbiographynophoto}{Ioannis Krikidis} (F’19) received the diploma in Computer Engineering from the Computer Engineering and Informatics Department (CEID) of the University of Patras, Greece, in 2000, and the M.Sc and Ph.D degrees from \'Ecole Nationale Sup\'erieure des T\'el\'ecommunications (ENST), Paris, France, in 2001 and 2005, respectively. He is currently a Professor at the Department of Electrical and Computer Engineering, University of Cyprus, Nicosia. His current research interests include wireless communications, 6G, wireless powered communications, RIS, quantum information processing.
\end{IEEEbiographynophoto}
\vspace{-1.2cm}
\begin{IEEEbiographynophoto}{Valentin Gilbert} received the diploma in Computer Engineering from the Computer Engineering and Informatics Department of the University of Technology of Troyes, France, in 2019, and the Ph.D degree from Universit\'e Paris-Saclay, Paris, France, in 2025. He is currently the founder and senior researcher at Qounselor.
\end{IEEEbiographynophoto}

\end{document}